\documentclass[english,prl,twocolumn]{revtex4}
\usepackage[T1]{fontenc}
\usepackage[latin1]{inputenc}
\usepackage{graphicx}
\usepackage{amsmath}
\usepackage{aas_macros}
\usepackage{color}

\makeatletter
\usepackage{babel}
\makeatother
\begin{document}

\title{Non-commuting Observables are Jointly Measureable  under Disturbance Correction Strategy}

\author{Liang-Liang Sun, Yong-Shun Song, Zhi-Xin Chen and Cong-Feng Qiao}
\affiliation{Department of Physics,
University of Chinese Academy of Sciences, YuQuan Road 19A, Beijing 100049, China}
\date{\today{}}

\begin{abstract}
 In the study of Heisenberg's error-disturbance relation, it is commonly believed that the non-unitary change of states hinders us from deducing the information  encoded in original states about subsequently measured observable. However, we find that the disturbance can be corrected iff  the pre-measurement is non-projective. In this work, by analysing the effect of  decoherence on statistics of the subsequential measurement, we find the acquired information from pre-measurement can be used to developed a correction strategy,   and then the information about  post-measured observable can be recovered. In viewpoint of estimation theory, this result is the unbiasedness condition, which enable us to define precisions of measurement directly in terms of Fisher information. Moreover, we study the precisions trade-off relations in the information theoretic viewpoint.

\end{abstract}


\maketitle

\section{I Introduction}
\label{sec:intro}
Heisenberg originally formulated a trade-off relation for error of a position measurement and the resulting disturbance to subsequent measurement of momentum \cite{heisenberg,H,K}. In fact, the relations deal with the intrinsic uncertainty any quantum state must possess, not the precision of a measurement and the disturbance it introduces.

Quantum measurement is fundamentally statistical \cite{RMP}, the information about the observables is interpreted as the expectation value \cite{11,12} or statistics of a projective measurement\cite{19}. In view of this, decorhence occurs when a measurement is performed on states, affecting the statistics of a subsequent measurement. it has been widely believed that we cannot retrieve information encoded in original state about the incompatible observables \cite{11,kr0,oza} from a sequential measurements. Recently, in order to capture the spirit of precision limit by Heisenberg, many Measurement-Disturbance-Relations (MDRs) were proposed, meanwhile many controversies around the MDRs arise due to disagreement about proper definitions of error and disturbance.

Notwithstanding the attention paid on MDRs \cite{15,16,17,18}, there is little discussion contributing to the understanding of how the  decoherence affect the the statistics of a subsequent measurement. Note also if we want to fully  quantify the precision limit of joint measurements, we have to take any possible remedy aimed at optimizing the performance of the measurement outcomes. A question may arise as to whether there exist a "correction" strategy for the disturbed information allow us to acquire the informations encoded in the original states  about the non-commute observables, if so, how we are going understand the measuring precisions of the non-commuting observables.

In this work, we try to clarify the subject of "disturbance" by dividing the disturbed probabilities from post-measurement into decoherence-dependent parts and decoherence-independent parts. We shall show that decoherence-dependent parts are subject to a diminution by a decoherence related factor, and the decoherence-independent parts are not affected.  Moreover it will be seen that this diminution can be remedied iff the pre-measurement is non-projective, which can be interpreted as that the complete information about incompatible observables can be extracted from joint measurements \cite{13,14}.  In estimation theory,  which is the unbiasedness condition.

Under the correction strategy, the estimation error \cite{15,16,17,18} is corrected, so the only difference between independent projective measurements and joint measurements is the amount of information extractable from every sample. In an information-theoretic viewpoint, Fisher information (FI) \cite{4} and quantum Fisher information (QFI) \cite{7,8,YU book,6} are key notions in quantifying the amount of information that one can extracted \cite{333} from observed probability distributions. FI is sufficient for our following investigation \cite{zhu}. As unbiasedness condition is required in definition of FI, and lacking generality of this condition \cite{JD,11,12,17} has ever limited the investigation about uncertainty relation in this powerful tool \cite{YU book}.  By our correction strategy, unbiasedness condition achieves generality, and the precisions can be well defined in terms of FI,  then which can be studied in an information theoretic viewpoint.

Our paper is outlined as follows. In section II, we analysis the original of disturbance and introduce the correction strategy. In section III, we define precisions of measuring $\hat{A}$ and measuring $\hat{B}$  in term of the Fisher information, and a trade-off precisions relation is showed in a whole range measurement strength.
\section{II Joint measurability with correction strategy}
Quantum measurement is fundamentally statistical, the informations of the measured observables are interpreted as statistics of a projective measurement of the observables. Ideal measurements definitely should enable us to acquire the information in a most informative way \cite{11,YU book}. In this paper, general quantum measurement processes are investigated in terms of information completeness and "informativity".

\emph{\textbf{Definition}}. (Information completeness and Unbiasedness condition) \emph{If statistics of a perfect measurement can be deduced from the obtained probabilities without any prior information about the measured states, we say the information obtained is complete, or equivalently, estimator $\varphi^{A}_{est}$ of $\langle \hat{A}\rangle$ can be constructed with observed quantifies, so that $|\varphi^{A}_{est}-\langle \hat{A}\rangle|$ is close to zero with the independent repetitions of the measurement, meanwhile $\varphi^{A}_{est}$ is unbiased estimator.}

However, measurement on states quite likely affects the statistics of a subsequent measurement on the states, which hinders us from deducing the information about the post-measured observable encoded in original state. However, if we regard the information obtained in pre-measurement as prior information to the subsequent measurement, we can greatly optimize the performance of the post-measurement outcomes.

Here, we focus on the measurements in a qubit Hilbert space  the measured state is
$\mid\Phi\rangle=\sin\alpha|0\rangle_{s}+\cos\alpha e^{i\phi}|1\rangle_{s}$,  $\hat{A}=\sigma_{z}$ and $\hat{B}=\overrightarrow{\sigma}\cdot\overrightarrow{n}$ are the incompatible observables concerned, where $\overrightarrow{\sigma}=(\sigma_{x},\sigma_{y},\sigma_{z})$, $\sigma_{i}$ are Pauli matrixs, $\overrightarrow{n}=(\sin\theta\cos\varphi,\sin\theta\sin\varphi,\cos\theta)$,  $|0\rangle,|1\rangle$ and $|+\rangle,|-\rangle$ are two sets eigenvectors of $\hat{A}$ and $\hat{B}$ with eigenvalue 1 and -1.  The independent probability in projective measurement of $\hat{A}$ is $\langle\Phi|0\rangle\langle0|\Phi\rangle=\sin^{2}\alpha$, and $\sin^{2}\alpha$ can be used to characterize the information of  $\hat{A}$, note also the relative phase $\phi$ is redundant for measuring $\hat{A}$. Similarily, $\langle\Phi|+\rangle\langle+|\Phi\rangle$ can be used to characterize the information we want to extract from the post-measurement, where $\langle\Phi|+\rangle\langle+|\Phi\rangle=\sin^{2}\alpha\cos^{2}\frac{\theta}{2}+\cos^{2}\alpha\sin^{2}\frac{\theta}{2}+\sin2\alpha\sin\frac{\theta}{2}\cos\frac{\theta}{2}\cos(\varphi-\phi)$. Interestingly, it can be seen that the phase is essential for measuring $\hat{B}$.
 
 Consider the following sequential measurement scenario employed in \cite{13,14,16,45}, the measurements are realized by "weakly" coupling the signal photons to meter photons which is then subject to projective measurements, the "weakly" means that the states of signal photons are not completely decohered after the projective measurement rather than the change of state can be omitted.  And then, projective measurements of $\hat{B}$ are performed on the partial decohered signal photons states. Meter photons and signal photons are denoted by $m$, $s$.

The estimated informations about sequential measured observables are required to satisfy the following condition if the joint measurements can produce complete information:
 \begin{eqnarray}
F_{a_{i}}(P_{m=1},P_{m=-1})=tr(|\Phi\rangle\langle\Phi|i_{A}\rangle\langle i_{A}|),\\
F_{b_{i}}(P_{m=1},P_{m=-1},P_{B_{\eta}=1},P_{B_{\eta}=1})=tr(|\Phi\rangle\langle\Phi|i_{B}\rangle\langle i_{B}|).
\end{eqnarray}
Where $P_{m=i},P_{B_{\eta}=i}$ are the probabilities obtained from the joint measurements, $|i_{A}\rangle, |i_{B}\rangle$ are eigenvectors. $F_{i}$ are functions of the probabilities, which rely on measurement arrangement. $P_{m=i}$ being taken account in the subsequent estimation implies that the remedy for disturbance is employed.

After signal and meter photons interact before either are measured,  the states of system in two-qubit subspace are:
 \begin{eqnarray}\nonumber
|\Psi\rangle=(\sin\alpha\gamma|0\rangle_{s}+\cos\alpha e^{i\phi}\overline{\gamma}|1\rangle_{s})|0\rangle_{m}+\\
+(\sin\alpha\overline{\gamma}|0\rangle_{s}+\cos\alpha e^{i\phi}\gamma|1\rangle_{s})|1\rangle_{m}.
\end{eqnarray}
where $\gamma^{2}+\overline{\gamma}^{2}=1$,$\gamma$ and $\bar{\gamma}$  are real coupling parameters can be set through arranging measurement parameters. Without loss of generality, we take$\gamma\in(\frac{\sqrt{2}}{2},1)$.
The measurement strength is given as \cite{16}:
 \begin{eqnarray}
 \kappa=2\gamma^{2}-1.
\end{eqnarray}
The information about $\hat{A}$ is extracted by projective measuring the meter phontons on basis $|1\rangle_{m}, |0\rangle_{m}$, probabilities are obtained as:
\begin{eqnarray}\nonumber
p_{m=1}&=&\gamma^{2}\sin^{2}\alpha+{\overline{\gamma}}^{2}\cos^{2}\alpha\\
&=&\kappa\sin^{2}\alpha+\overline{\gamma}^{2},\\\nonumber
p_{m=-1}&=&\gamma^{2}\cos^{2}\alpha+{\overline{\gamma}}^{2}\sin^{2}\alpha\\
&=&\kappa\cos^{2}\alpha+\overline{\gamma}^{2}.
\end{eqnarray}
 Roughly speaking, the amount of target information $\sin^{2}\alpha$ encoded in the (5) and (6) is  directly proportional to the measurement strength $\kappa$, an accurate method for quantifying the amount will be introduced in next section. Based on probability distribution obtained, the information of $\hat{A}$ is given as:
\begin{eqnarray}
\langle\Phi|0\rangle\langle0|\Phi\rangle&=&\frac{p_{m=1}-\overline{\gamma}^{2}}{\kappa},\\
\langle\Phi|1\rangle\langle1|\Phi\rangle&=&\frac{p_{m=-1}-\overline{\gamma}^{2}}{\kappa}.
\end{eqnarray}
And the unbiased estimator of expectation value is given as:
\begin{eqnarray}
\varphi^{A_{\epsilon}}_{est}=\frac{p_{m=1}-p_{m=-1}}{\kappa}.
\end{eqnarray}

The states after the measurement can be calculated by taking a partial trace over the meter photons, giving us the mixed states corresponding to the system:
\begin{eqnarray}\nonumber
\rho_{s_{\gamma}}&=&tr_{m}(|\Psi\rangle\langle\Psi|)\\
&=&\left(
  \begin{array}{cc}
  \cos^{2}\alpha & 2\gamma\overline{\gamma} \sin\alpha\cos\alpha e^{-i\phi} \\
  2\gamma\overline{\gamma}\sin\alpha\cos\alpha e^{i\phi} & \sin^{2}\alpha \\
  \end{array}
  \right).
\end{eqnarray}
The degree to which the states decohered are corresponding to the $2\gamma\overline{\gamma}$.  When $\gamma\approx\frac{1}{\sqrt{2}}$ or $\kappa\approx0$ , no information is obtained, and the states are not affected. When $\gamma\approx1$ or $\kappa\approx1$ ,  projective measurements are performed, the information encoded in the obtained probabilities distribution is maximum, but the anti-diagonal terms in the matrix disappear, which contain the information of phase, which is essential for measuring  $\hat{B}$.  In a general weak measurement scheme, $2\gamma\overline{\gamma}$ is always smaller than $1$, and then the states partial decohered, we will see that the states carry a "obscure" information about about $\hat{B}$.

Subsequently, projective measurement of $\hat{B}$ is performed on basis $|+\rangle$ and $|-\rangle$. Probabilities are given as:
\begin{eqnarray}\nonumber
p_{B_{\eta}=1}=tr(\rho_{s_{\gamma}}|+\rangle\langle+|)=\sin^{2}\alpha\cos^{2}\frac{\theta}{2}\\\nonumber
+\cos^{2}\alpha\sin^{2}\frac{\theta}{2}+\frac{1}{2}(2\gamma\overline{\gamma})\sin2\alpha\sin\theta\cos(\varphi-\phi)\\
=(1-2\gamma\overline{\gamma})n+2\gamma\overline{\gamma}\langle\Phi|+\rangle\langle+|\Phi\rangle,\\\nonumber
p_{B_{\eta}=-1}=tr(\rho_{s_{\gamma}}|-\rangle\langle-|)=\cos^{2}\alpha\cos^{2}\frac{\theta}{2}\\\nonumber
+\sin^{2}\alpha\sin^{2}\frac{\theta}{2}-(\frac{1}{2})2\gamma\overline{\gamma}\sin2\alpha\sin\theta\cos(\varphi-\phi)\\
=(1-2\gamma\overline{\gamma})(1-n)+2\gamma\overline{\gamma}\langle\Phi|-\rangle\langle-|\Phi\rangle,
\end{eqnarray}
where we see the probabilities can be devided into two parts. For $p_{B_{\eta}=1}$, the first part is $\sin^{2}\alpha\cos^{2}\frac{\theta}{2}+\cos^{2}\alpha\sin^{2}\frac{\theta}{2}$, it is independent on the pre-measurement strength and not affected. The other part is $\sin2\alpha\sin\frac{\theta}{2}\cos\frac{\theta}{2}\cos(\varphi-\phi)$,  which is directly related to off-diagonal terms of the density matrix, being diminished by a decoherence related factor $2\gamma\overline{\gamma}$, which is the very core of disturbance. Note $2\gamma\overline{\gamma}\langle\Phi|+\rangle\langle+|\Phi\rangle$ can be regarded as a diminished target information.

In order to derive the information encoded in the original states from the disturbed probabilities, we introduce correction strategy. Firstly, with the $\sin^{2}\alpha$ obtained from pre-measurement outcomes (5) and (6), we construct the decoherence-independent parts, which enable us to extract the decoherence-independent part from (11). Secondly, the  information can be constructed  with the decoherence-independent part and the decoherence-dependent part. Though a easy calculation, the information encoded in original states is given as:
\begin{eqnarray}\nonumber
\langle\Phi|+\rangle\langle+|\Phi\rangle=\frac{1}{\kappa(1-2\gamma\overline{\gamma})}\times(\kappa p_{B_{\eta}=1}\\
-(1-2\gamma\overline{\gamma})(\cos^{2}\frac{\theta}{2}p_{m=1}+\sin^{2}\frac{\theta}{2}p_{m=-1}-\overline{\gamma}^{2})),\\\nonumber
\langle\Phi|-\rangle\langle-|\Phi\rangle=\frac{1}{\kappa(1-2\gamma\overline{\gamma})}\times(\kappa p_{B_{\eta}=-1}\\
-(1-2\gamma\overline{\gamma})(\sin^{2}\frac{\theta}{2}p_{m=1}+\cos^{2}\frac{\theta}{2}p_{m=-1}-\overline{\gamma}^{2})).
\end{eqnarray}
And the unbiased estimator of $\langle\Phi|\hat{B}|\Phi\rangle$ can be given as:
\begin{eqnarray}
\varphi^{B_{\eta}}_{est}=\frac{p_{B_{\eta}=1}-p_{B_{\eta}=-1}-(1-2\gamma\overline{\gamma})\cos\theta\varphi^{A_{\epsilon}}_{est}}{2\gamma\bar{\gamma}}.
\end{eqnarray}
In the above discussion, only $\gamma\neq1,\frac{1}{\sqrt{2}}$  are needed to guarantee information completeness or unbiasedness condition for the "incompatible" measurements. This is in strong contrast to the widely belief that the pre-measurement needed to be so weak that the disturbance to the state can be neglected\cite{oza,16}. We realised that pre-measurement can give prior information as well as introduce disturbance to subsequential measurement, which can be used to develop a correction for the disturbance.

Here, we are also interested in the relation between the decoherence-dependent part $\sin2\alpha\sin\theta\cos(\varphi-\phi)$ and the commutator $|\langle \hat{A}\hat{B}-\hat{B}\hat{A}\rangle| = 2\sin2\alpha\sin\theta\sin(\varphi-\phi)$. They are similar in the formula, implying that the degree to which the pre-measurement affect the statistics of a subsequent measurement is corresponding to but not fully in accordance with non-commutativity figure of the measured observables. This observation can result necessary and sufficient condition for zero-noise, zero-disturbance (ZNZD) states proposed in \cite{19}. The main idea in the Ref\cite{19} is that, in any d-dimensional Hilbert space and for any pair of non-commuting operators, $\hat{A}$ and $\hat{B}$, there exists at least $2^{d-1}$ zero-noise, zero-disturbance (ZNZD) states, measurement on the states that does not affect the statistics of a subsequent measurement. We find "the statistics is not affected" is equivalent to that the decoherence-dependent equal to zero. When $\cos(\varphi-\phi)=0$ and $\sin2\alpha\neq0$  $\sin\theta\neq0$ the state is a non-trivial ZNZD state, simultaneously there is no limit on $\alpha$, so there are infinite non-trivial ZNZD states.

\section{III Defining Precisions and tradeoff relations}
Recently, there is lively debate on MDRs, most of the controversies around the relations arise due to disagreement about proper definitions of error and disturbance(a detailed analysis of limitions of most common used definitions can be found in \cite{60}). Fisher information is key quantity to characterize the ultimate precision in parameter estimation. However, the requirement of unbiasedness condition has limited the use of FI in investigation in uncertainty relation\cite(YU book). Because of the correction strategy, we needn't worry about that.

Regarding the pre-measurement, the stronger the measured states interact with the measurement devices, the more information can be extracted from (5) and (6), and the less information available from (11) and (12). Therefore, this analysis should yield a precisions trade-off relation. In consideration of informativity, we refer to amount of information extractable from one experiment record \cite{4,8,YU book,22}, which can be quantified by FI \cite{333}. Formally, FI is the expectation value of the observed information.
\begin{eqnarray}
I=\sum_{i}P_{i,\theta}(\frac{\partial\ln P_{i,\theta}}{\partial\theta})^{2}.
\end{eqnarray}
Where $P_{i,\theta}$ is the probabilities used to extracted information about $\theta$.
Here we define the FI for the joint measurement process. Which is amount of the information about $\hat{A}$ and $\hat{B}$ that one time joint experiment records contained. $P_{m=i},P_{B_{\eta}=i}$ is probabilities corresponding to the joint measurements:
\begin{eqnarray}\nonumber
I^{A}_{\epsilon}&=&\sum_{i}P_{m=i}(\frac{\partial\ln P_{m=i}}{\partial\langle\Phi|\hat{A}|\Phi\rangle})^{2}\\
&=&\frac{1}{4}\frac{(1-2\gamma^{2})^{2}}{(p_{m=1})(p_{m=-1})},\\\nonumber
I^{B}_{\eta}&=&\sum_{i}P_{B_{\eta}=i}(\frac{\partial\ln P_{B_{\eta}=i}}{\partial\langle\Phi|\hat{B}|\Phi\rangle})^{2}\\
&=&\frac{1}{4}\frac{(2\gamma\overline{\gamma})^{2}}{(p_{B_{\eta}=1})(p_{B_{\eta}=-1})}.
\end{eqnarray}
Similarly, Fisher information for independent projective measurement of $\hat{A}$ and $\hat{B}$ can be defined as:
 \begin{eqnarray}\nonumber
I^{A}&=&\sum_{i}(\frac{\partial\ln p_{A=i}}{\partial\langle\Phi|\hat{A}|\Phi\rangle})^{2}p_{A=i}\\
      &=&\frac{1}{4}\frac{1}{(p_{A=1})(p_{A=-1})},
\end{eqnarray}
\begin{eqnarray}\nonumber
I^{B}&=&\sum_{i}(\frac{\partial\ln p_{B=i}}{\partial\langle\Phi|B|\Phi\rangle})^{2}p_{B=i}\\
     &=&\frac{1}{4}\frac{1}{(p_{B=1})(p_{B=-1})}.
\end{eqnarray}
where $p_{A=1}=tr(|\Phi\rangle\langle\Phi||0\rangle\langle0|)$, $p_{A=-1}=tr(|\Phi\rangle\langle\Phi||1\rangle\langle1|)$,  $p_{B=1}=tr(|\Phi\rangle\langle\Phi||+\rangle\langle+|)$,$ p_{B=-1}=tr(|\Phi\rangle\langle\Phi||-\rangle\langle-|)$.

The precision to which we can estimate the parameter is fundamentally limited by the FI, for only finite $n$ ensembles are available in laboratory,  statistic errors of the estimated expectation value is given by the Cram$\acute{e}$r-Rao inequality \cite{8}:
\begin{eqnarray}
Var_{n}[\varphi_{n}]\geq\frac{1}{nI}.
\end{eqnarray}

Comparing independent projective measurements, a very natural definition of precisions can be given as follow:
\begin{eqnarray}
\epsilon&=&\frac{I_{\epsilon}^{A}}{I_{A}},\\
\eta&=&\frac{I_{\eta}^{B}}{I_{B}}.\\\nonumber
\end{eqnarray}
Which are the rations of amount of information gained from one time joint measurement records to that from a projective measurement record.
In contrast with errors and disturbances \cite{15,16,17,18,JD,11,kr0,oza} defined before, the precisions are directly defined, moreover estimation errors in the Refs are eradicated from the disturbed information, which allow complete information about the non-commuting observables to be extractable.  and hence our definition of precisions yield precise mathematical translations of Heisenberg's idea about precision limit. In fig.1, the trade-off relation about the precisions of incompatible measurements is showed.
\begin{figure}
\begin{center}
\includegraphics[width=0.35\textwidth]{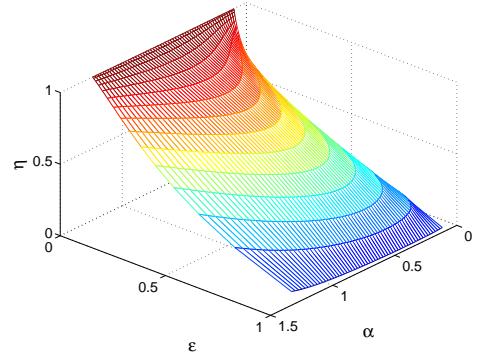}
\end{center}
\vspace{+0.9cm}
\caption{The precision of post-measurement $\eta$ decrease with the increasing of  pre-measurement precision $\epsilon$, till to zero when the pre-measurement is projective. The precisions of the pre-measurement and post-measurement concerned are mainly determined by the measurement arrangement, depend on the state little, this is a very good feature. Where the figure is showed when $\theta=\frac{\pi}{3}, \phi-\varphi=0$. $\alpha$ reflect partial information about the measured states} \label{fig2}
\end{figure}

\section{Conclusion}
In summery, we demonstrate how decoherence due to pre-measurement affect the subsequential observed statistics, and a correction strategy for the disturbed information is developed. Then we show that the informations about the non-commute observables encoded in original state are extractable iff the pre-measurement is non-projective. This result is in strong contrast with the universally regarded that the pre-measurement needed to be so weak that the change of state can be omitted. Furthermore,  in viewpoint of estimation theory, the result is the unbiasedness condition, which enable us to define precisions of measurement directly in terms of Fisher information, settling the dispute over the descriptions of precisions. The precisions trade-off relations are also showed. Our results offer fundamental insights on understanding of quantum complementarity and uncertainty principle,  Note also our correction strategy produce unbiased estimation of the parameter, we believe which will spur the investigation of the application of joint measurements schemes in state estimation and quantum metrology.


\begin{thebibliography}{99}
\bibitem{heisenberg}
W. Hersenberg, Zeit. f$\ddot{u}$r Physik {\bf 43}, 172 (1927).
\bibitem{H} H. P. Robertson, Phys. {\bf34}, 163(1929).
\bibitem{K} E. H. Kennard, Z.Phys. {\bf44}, 326 (1927).
\bibitem{RMP} A. Peres, D. R. Terno, Rev. Mod. Phys. {\bf76}, 93  (2004).
\bibitem{11} Y. Watanabe, T. Sagawa, M. Ueda,  Phys. Rev. A {\bf84}, 042121 (2011).
\bibitem{12} Y. Watanabe, M. Ueda, arXiv:1106.2526 (2011).
\bibitem{19} K. Korzekwa, D.Jennings, and T. Rudolph, arxiv:1311.5506V2.
\bibitem{kr0} H. T. Lim, Y. S. Ra, K. H. Hong, S. W. Lee, Y. H. Kim, Phys. Rev. Lett. {\bf113}, 020504 (2014).
\bibitem{oza} F. Buscemi, M. J. W. Hall, M. Ozawa, and M. M. Wilde, Phys. Rev. Lett. {\bf112}, 050401 (2014).
\bibitem{15} A. Lund and H. M. Wiseman, New J. Phys. {\bf12}, 093011(2010).
\bibitem{16} G. J. Pryde, J. L. O$'$Brien, A. G. White, T. C. Ralph, and H. M. Wiseman Phys. Rev. Lett. {\bf94}, 220405 (2005).
\bibitem{17} M. Ozawa, Phy. Rev. A {\bf67}, 042105 (2003).
\bibitem{18} C. Branciard, arXiv:1312.1857.
\bibitem{13} Y. Aharonov, D. Z. Albert, and L. Vaidman, Phys. Rev. Lett.  {\bf60}, 1351 (1988).
\bibitem{14} J. S. Lundeen and A. M. Steinberg, Phys. Rev. Lett. {\bf102}, 020404.
\bibitem{4} O. E. Barndorff-Nielsen R. D. Gill Arxiv:9808009V4.
\bibitem{7}  D. Petz, C. Ghinea, arXiv:1008.2417.
\bibitem{8} $\acute{A}$ngel Rivas, A. Luis, Phys. Rev. Lett. {\bf105}, 010403 (2010).
\bibitem{YU book} Y. Watanabe, [Formulation of Uncertainty Relation Between Error and Disturbance in Quantum Measurementby Using Quantum Estimation Theory], ISSN 2190-5053.
\bibitem{333} Lehmann, E. L. Casella, G. (1998). Theory of Point Estimation (2nd ed.). Springer. ISBN 0-387-98502-6.
\bibitem{6}  S. Luo. Phys. Rev. Lett. {\bf91}, 180403 (2003).

\bibitem{zhu}  Actually, we deduce information about observables from classic probabilities obtained in laboratory, so FI is sufficient.
\bibitem{JD} J. Dressel, F. Nori, Phys. Rev. A {\bf89}, 022106 (2014).
\bibitem{45} L. A. Rozema, A. Darabi, D. H. Mahler, A. Hayat, Y. Soudagar, and A. M. Steinberg, Phys. Rev. Lett. {\bf109}, 100404  (2012).


\bibitem{60}  P. Busch, T. Heinonen, Physics Letters A. {\bf320} 261 (2010).
\bibitem{22}  P. Facchi, R. Kulkarni, V. I. Man'ko, G. Marmo, E. C. G. Sudarshan, F. Ventriglia,  Physics Letters A. {\bf374}  4801 (2010).







\end{thebibliography}
\end{document}